\documentclass[epsfig]{article}
\usepackage[iso-8859-7]{inputenc}
\usepackage{epsfig}
\usepackage{amsmath}
\catcode`\@=11

\def\@normalsize{\@setsize\normalsize{15pt}\xiipt\@xiipt
\abovedisplayskip 14pt plus3pt minus3pt%
\belowdisplayskip \abovedisplayskip
\abovedisplayshortskip  \z@ plus3pt%
\belowdisplayshortskip  7pt plus3.5pt minus0pt}
\def\small{\@setsize\small{13.6pt}\xipt\@xipt
\abovedisplayskip 13pt plus3pt minus3pt%
\belowdisplayskip \abovedisplayskip
\abovedisplayshortskip  \z@ plus3pt%
\belowdisplayshortskip  7pt plus3.5pt minus0pt
\def\@listi{\parsep 4.5pt plus 2pt minus 1pt
            \itemsep \parsep
            \topsep 9pt plus 3pt minus 3pt}}

\def\underline#1{\relax\ifmmode\@@underline#1\else
        $\@@underline{\hbox{#1}}$\relax\fi}
\@twosidetrue \relax

\catcode`@=12

\catcode`\@=11

\def\section{\@startsection{section}{1}{\z@}{3.5ex plus 1ex minus
   .2ex}{2.3ex plus .2ex}{\large\bf}}

\def\ps@headings{\def\@oddfoot{}\def\@evenfoot{}
\def\@oddhead{\hbox{}\hfill
        \makebox[.5\textwidth]{\raggedright\ignorespaces --\thepage{}--
        \hfill }}
\def\@evenhead{\@oddhead}
\def\subsectionmark##1{\markboth{##1}{}}
}

\ps@headings

\catcode`\@=12

\relax

\def\figcap{\section*{Figure Captions\markboth
        {FIGURECAPTIONS}{FIGURECAPTIONS}}\list
        {Fig. \arabic{enumi}:\hfill}{\settowidth\labelwidth{Fig. 999:}
        \leftmargin\labelwidth
        \advance\leftmargin\labelsep\usecounter{enumi}}}
 \relax
\def\tablecap{\section*{Table Captions\markboth
        {TABLECAPTIONS}{TABLECAPTIONS}}\list
        {Table \arabic{enumi}:\hfill}{\settowidth\labelwidth{Table 999:}
        \leftmargin\labelwidth
        \advance\leftmargin\labelsep\usecounter{enumi}}}
 \relax
\def\reflist{\section*{References\markboth
        {REFLIST}{REFLIST}}\list
        {[\arabic{enumi}]\hfill}{\settowidth\labelwidth{[999]}
        \leftmargin\labelwidth
        \advance\leftmargin\labelsep\usecounter{enumi}}}
 \relax

\catcode`\@=11

\def\marginnote#1{}
\newcount\hour
\newcount\minute
\newtoks\amorpm
\hour=\time\divide\hour by60 \minute=\time{\multiply\hour by60
\global\advance\minute by- \hour}
\edef\standardtime{{\ifnum\hour<12 \global\amorpm={am}%
    \else\global\amorpm={pm}\advance\hour by-12 \fi
    \ifnum\hour=0 \hour=12 \fi
    \number\hour:\ifnum\minute<100\fi\number\minute\the\amorpm}}
\edef\militarytime{\number\hour:\ifnum\minute<100\fi\number\minute}

\def\draftlabel#1{{\@bsphack\if@filesw {\let\thepage\relax
  \xdef\@gtempa{\write\@auxout{\string
    \newlabel{#1}{{\@currentlabel}{\thepage}}}}}\@gtempa
    \if@nobreak \ifvmode\nobreak\fi\fi\fi\@esphack}
     \gdef\@eqnlabel{#1}}
\def\@eqnlabel{}
\def\@vacuum{}
\def\draftmarginnote#1{\marginpar{\raggedright\scriptsize\tt#1}}
\def\draft{\oddsidemargin -.5truein
        \def\@oddfoot{\sl preliminary draft \hfil
        \rm\thepage\hfil\sl\today\quad\militarytime}
        \let\@evenfoot\@oddfoot \overfullrule 3pt
        \let\label=\draftlabel
        \let\marginnote=\draftmarginnote

\def\@eqnnum{(\theequation)\rlap{\kern\marginparsep\tt\@eqnlabel}%
\global\let\@eqnlabel\@vacuum}  }
\def\preprint{\twocolumn\sloppy\flushbottom\parindent 1em
        \leftmargini 2em\leftmarginv .5em\leftmarginvi .5em
        \oddsidemargin -.5in    \evensidemargin -.5in
        \columnsep 15mm \footheight 0pt
        \textwidth 250mmin      \topmargin  -.4in
        \headheight 12pt \topskip .4in
        \textheight 175mm
        \footskip 0pt

\def\@oddhead{\thepage\hfil\addtocounter{page}{1}\thepage}
        \let\@evenhead\@oddhead \def\@oddfoot{} \def\@evenfoot{}
}
\def\titlepage{\@restonecolfalse\if@twocolumn\@restonecoltrue\onecolumn
     \else \newpage \fi \thispagestyle{empty}\c@page\z@
        \def\thefootnote{\fnsymbol{footnote}} }
\def\endtitlepage{\if@restonecol\twocolumn \else  \fi
        \def\thefootnote{\arabic{footnote}}
        \setcounter{footnote}{0}}  
\catcode`@=12 \relax

\def\ps@headings{\def\@oddfoot{}\def\@evenfoot{}
\def\@oddhead{\hbox{}\hfill
        \makebox[.5\textwidth]{\raggedright\ignorespaces --\thepage{}--
        \hfill }}
\def\@evenhead{\@oddhead}
\def\subsectionmark##1{\markboth{##1}{}}
}

\ps@headings

\relax

\def\firstpage#1#2#3#4#5#6{
\begin{document}
\newcommand{\newc}{\newcommand}
\newc{\ra}{\rightarrow}
\newc{\lra}{\leftrightarrow}
\newc{\beq}{\begin{equation}}
\newc{\be}{\begin{equation}}
\newc{\eeq}{\end{equation}}
\newc{\ee}{\end{equation}}
\newc{\ba}{\begin{eqnarray}}
\newc{\ea}{\end{eqnarray}}
\def\eps{\epsilon}
\def\la{\lambda}
\def\f{\frac}
\def\nm{\nu_{\mu}}
\def\nt{\nu_{\tau}}
\def\sq{\sqrt{2}}
\def\ri{\rightarrow}
\newc{\sm}{Standard Model}
\newc{\smd}{Standard Model}
\newc{\barr}{\begin{eqnarray}}
 \newc{\earr}{\end{eqnarray}}
\begin{titlepage}
\nopagebreak
\title{\begin{flushright}
        \vspace*{-0.8in}
{
}
\end{flushright}
\vfill {#3}}
\author{\large #4 \\[1.0cm] #5}
\maketitle \vskip -7mm \nopagebreak
\begin{abstract}
{\noindent #6}
\end{abstract}
\vfill
\begin{flushleft}
\rule{16.1cm}{0.2mm}\\[-3mm]

\end{flushleft}
\thispagestyle{empty}
\end{titlepage}}

\def\simlt{\stackrel{<}{{}_\sim}}
\def\simgt{\stackrel{>}{{}_\sim}}
\date{}
\firstpage{3118}{IC/95/34} {\large\bf Gravitational Atom in
Compactified Extra Dimensions} {E.G. Floratos$^{\,a,\,b}$, G.K.
Leontaris$^{\,c}$
and N.D. Vlachos$^{\,d}$}
{\normalsize\sl
$^a$Physics Deptartment, University of Athens,
 Zografou 157 84 Athens, Greece \\[2.5mm]
 \normalsize\sl $^b$ Institute of Nuclear Physics, NCSR Demokritos, 15310, Athens, Greece
 \\[2.5mm]
\normalsize\sl $^c$Theoretical Physics Division, Ioannina
University,
GR-45110 Ioannina, Greece\\[2.5mm]
\normalsize\sl
$^d$Theoretical Physics Division, Aristotle University, GR-54006 Thessaloniki, Greece
 \\[2.5mm]
 }
{ We consider quantum mechanical effects of the modified  Newtonian potential in the presence of extra compactified  dimensions. We develop a method to solve the resulting Schr\"odinger equation and determine the energy shifts caused by the Yukawa-type corrections of the potential. We comment on the possibility of detecting the modified gravitational bound state Energy spectrum by  present day and future experiments.
}

\newpage

\section{Introduction}

Over the last few decades considerable experimental work has been  devoted to test the accuracy of Newton's Gravitational Inverse Square Law (ISL)
at short distances. To that end, a number of experiments  using various sophisticated devices were designed to test the validity of ISL in distances as
small as the sub-millimeter scale.  Nowadays, one of the main theoretical motivations  stimulating these extensive experimental searches is the prediction
of Newton's Law modifications in theories with `large' extra dimensions. Indeed, String Theory  and related brane scenarios predict that our world is
immersed in a higher 10-dimensional space where six of the ten dimensions are compact. A particular class of string constructions~\cite{Antoniadis:1990ew,ArkaniHamed:1998rs} suggest that some of the extra dimensions could be decompactified at sub-millimeter distances,
and manifest themselves through modifications of gravity and, in particular, of the Inverse Square Law. Recent experiments have tested  the validity of  ISL
down to the scale of a few microns depending on the particular model and the experimental methodology~\cite{Long:1998dk}-\cite{Decca:2009fg}.

Another class of experiments that have also been revived today measure quantum gravitational  effects \cite{van der Zouw:2000ay}-\cite{Nesvizhevsky:2004qb}~\footnote{For recent experimental status  see\cite{Antoniadis:2007zz}.}.
In order to avoid the dominance of electromagnetic interactions  these experiments are performed with neutral particles. For instance, a neutron
interferometer to measure the quantum mechanical phase shift of neutrons due to the interaction  with   Earth's gravitational field was proposed a
long time ago in~\cite{Colella:1975dq}. In recent  experiments also, the quantum mechanical levels of a cold neutron  beam above  a flat optical
mirror in  Earth's gravitational field were also investigated~\cite{Nesvizhevsky:2002ef,Nesvizhevsky:2004qb}. According   to the predictions of models
with extra-dimensions, modifications to Newton's law increase  at shorter distances, in particular, close to or inside the compactification radius. Since
the scale of quantum mechanical effects is many orders of magnitude smaller than the sub-millimeter scale -which is the range probed  by the present
experiments-  possible modifications  could become very important and eventually measurable at the atomic level.

In the experiments, the common parametrization of the corrections to the Newton's potential is considered to be of Yukawa type. Thus, the total
potential is expressed as follows
\ba
\Phi(r)&=&-G_N\,\frac{M\,M'}r\,\left(1+\alpha\,e^{-r/\lambda}\right)\cdot
\label{CorrExp}
\ea
The parameter $\alpha$ characterizes the strength of the Yukawa type correction to gravity, while $\lambda$ accounts for the range of
this extra interaction term. A considerable number of experiments testing the Newtonian nature of gravity have  put strong
limits~\cite{Long:2003ta,Geraci:2008hb} on the strength and the range of the additional Yukawa interaction in (\ref{CorrExp}).

Remarkably,  it was found that the Yukawa-type correction in the above empirical formula (\ref{CorrExp}) is of the same form with the
leading correction term  of the potential derived in the presence of extra compact dimensions. In the case of toroidal compactification
in particular, it takes the form~\cite{Floratos:1999bv,Kehagias:1999my}
\ba
\Phi(r)&=&-G_N\,\frac{M\,M'}r\,\left(1+2\,n\,e^{-r/R_C}\right)
\label{ApprExp}
\ea
where $n$ is the number of extra dimensions and $R_C$ the compactification radius. The radius $R_C$ and the  effective Planck scale
$M_C$ can be related as follows~\cite{ArkaniHamed:1998rs}: The Gauss' Law for distances  $r\ll R$ results to the gravitational potential
\ba
\Phi(r)&=&-\left(\frac{\hbar}{c}\right)^{n}\frac{\hbar \,c}{M_C^{n+2}}\frac{M\,M'}{r^{n+1}}\
\cdot \label{smallr}
\ea
In the absence of extra dimensions, $n=0$ and $M_C=M_P$~\footnote{It is to be mentioned that the Planck mass  is expressed in terms
of  the gravitational constant as $M_P^{2}=\frac{\hbar c}{G_N }$.}  the above formula coincides with the standard four-dimensional
Newton's potential.

For distances much larger than the compactification radius, $r\gg R_C$, we should recover the Newton's potential, and the formula takes the
form
\ba
\Phi(r)&=&-\left(\frac{\hbar}{c}\right)^{n}\frac{\hbar\,c}{M_C^{n+2}R_C^n}\frac{M\,M'}{r}\,\cdot
\ea
Comparing with (\ref{ApprExp})
\ba
M_C^{n+2}R_C^n&=&\frac{\hbar^2}{G_N} \left(\frac{\hbar}{c}\right)^{n-1}
\ea
which implies the following numerical relation
\ba
R_C&=&\left(\frac{\hbar\,G_N}{c^3}\right)^{\frac 12}\left(\frac{M_{Pl}}{M_C}\right)^{1+\frac 2n}\;=\;\text{1.97}\times 10^{-17}
    e^{74.0821/n}
   \left(\frac{1\text{TeV}}{\text{$M_C$}}\right)^{1+\frac{2}{n}}\text{cm}\label{MCRC}
   \ea
Given the number $n$ of extra dimensions, formula (\ref{MCRC})
determines the radius as a function of the higher dimensional Planck
scale. Thus, for one extra compact dimension, $n=1$, a
string scale as low as $M_C\sim 10$TeV,  would lead to a `decompactified'
radius $R_C\sim 10^{10}$ meters, i.e. of the order of  solar
distances. For ranges up to this order, the Yukawa type correction
in (\ref{ApprExp}) is comparable to the  ordinary
gravitational term,  implying observable hard violations to Newton's
law. However, the scale $M_C$ is not
determined by some principle and can be anywhere between $M_W$ and
$M_P$.
It is observed that for $n=1$ and compactification scale less than
$\sim 10^9$ GeV, the compact radius is at most in the sub-millimeter
range, thus, at distances $r\sim R_C$ corrections become important
and would have been detected (for example, see relevant graphs in~\cite{Long:2003ta,Geraci:2008hb}).
 However, for $M_C\ge 10^{10}$GeV $R_C$ drops down to
$10^{-6}$cm. In the presence of more than one compact dimensions, it is
possible to considerably reduce  the compactification  scale without
contradicting the present day experiments. Thus, as it can be checked
from formula  (\ref{MCRC}) for $n=2$ and $M_C\sim 100$TeV  for example,
 we expect measurable modifications at distances $R\sim 10^{-7}$cm.
This should be compared, for example, with the Bohr radius
which is  defined as
\ba
a_0&=&\frac{4\pi\epsilon_0\hbar^2}{e^2\,m_e}\;=\; 5.29\times
10^{-9}{\rm cm}
\ea
The corresponding Bohr radius for a gravitational atom (gratom) containing  a
neutron instead of an electron could be defined as
\ba
a_G&=&\frac{\hbar^2}{G_N\,M_0\,m_n^2}\label{GA}
\ea
where $M_0$ is the mass generating the gravitational potential.

From the above discussion, we see that  experimental constraints
restrict the ${\lambda}\sim R_c$ radius  at minuscule
distances where quantum mechanical effects might be sizable. This way,
new experimental devices could possibly detect deviations from Newton's
law, or put more stringent bounds by means of appropriate quantum measurements.
 For example, in recent experiments, it has been shown that ultra-cold neutrons (UCN) in
the Earth's gravitational field form  bound states.
It turns out  that  consistency with  Newton's gravity is at the 10\%
level, so that bounds on non-standard gravity are put
at the nanometer scale~\cite{Nesvizhevsky:2004qb,Westphal:2007wm,Dimopoulos:2003mw}.

Motivated by the interesting results of the recent experimental activity,  in this letter,
we consider the quantum mechanical system  of a `gravitational atom' involving a
light  neutral elementary particle in the presence of  extra
compact dimensions. In particular,  we  study the corresponding
Schr\"odinger equation  that encodes the effects of the compact dimensions through a rather
 complicated modified Newton's potential, aiming to obtain
 the modifications on  measurable quantities~\footnote{Quantum mechanical effects from
 extra dimensions in various perspectives
 were studied also in references~\cite{Brandhuber:1999hb}.}.

\section{Gravitational Potential in the presence of extra compactified dimensions}

In this section, we review in brief the derivation of the  modified gravitational
potential implied by the existence of an arbitrary number of extra compact dimensions 
and analyze its behavior at various distances.
Then, we proceed to a mathematical analysis of the results and determine the
behavior of the potential at various distances with respect to the
 radii of the compactified extra dimensions.

 Let $\vec x=(x_1,x_2,x_3)$ and $\vec y=(y_1,y_2,y_3)$ be vectors of the
 the ordinary 3-dimensional space and $x_i,y_i$ their corresponding
  coordinates.
 Assuming toroidal compactification, we denote $x_i^c=R\,\theta_i$,
  $i=1,\dots,n$ the coordinates of the $n$ compact dimensions with
  $\theta_i=[0,2\pi)$ the corresponding angles while, for simplicity,
  we have adopted a common compactification radius $R_C$.
In the presence of $n$ compact extra dimensions the gravitational
potential for two unit masses obeys the Laplace equation\footnote{For
convenience, from now on we drop the index $C$ and simply write $R_C\ra R$.}
\ba
\nabla^2_{n+3}\Phi&=&-\mu\,\delta^3(\vec x-\vec
y)\frac{1}{R^n}\delta^n(\theta-\theta_0)
\ea
where  for simplicity we introduced the parameter $\mu$ to account for
various dimensional constants to be taken into account later on.
Using the Fourier transform and performing the appropriate
integrations in  momentum space and  restoring units,
the solution is found to be
\ba
\Phi(r,\theta)&=&-G_N\,\frac{M\,M'}r\,\left(1+2\,\sum_{\vec
m}^{\infty}e^{-\frac{|\vec{m}| r}{R}}\cos(\vec m\cdot\vec\theta)\right)\cdot
\label{pot}
\ea
where $r=|\vec x-\vec y|$, and the summation is over the tower of
KK-modes in the dimensions of the compact space  $\vec
m=(m_1,m_2,\dots, m_n)$.

 The first term in the potential  (\ref{pot}) generates the standard gravitational
inverse quare law for the induced force. The second term is an infinite sum on KK-modes
 due to the presence of extra dimensions and describes a short range interaction exponentially
suppressed by powers of $e^{-r/R}$. For distances much larger than
the compactification radius however, (i.e. for $r\gg R$), all the terms of this infinite
sum are highly suppressed by these exponential powers, thus (\ref{pot})
reduces to Newton's three dimensional analogue.

For  measurements in the vicinity of the
compactification radius $r\sim R$,  the behavior of the infinite sum
 is not manifest. In the case of one extra dimension however,
we may obtain an exact formula for the potential (\ref{pot}).
Setting $\vec m=m$ and performing the sum for $n=1$, we get
\ba
\Phi_{n=1}(r, \theta)&=&-G_N\,\frac{M\,M'}r\,\frac{e^{2r/R} -1}{e^{2 r/R}-2
e^{r/R} \cos \theta +1}\label{Vn1}\cdot
\ea
The  effect of the compact dimensions is maximized for $\theta=0$, where
for the $n=1$ case the potential assumes the simplified form
\ba
\Phi_{n=1}(r, 0)&=&-G_N\,\frac{M\,M'}r\,\coth\left(\frac{r}{2R}\right)\cdot
\label{pot01}
\ea
This formula  interpolates between large and small distances $r$
compared to the compactification scale $R_C$.

As already mentioned, for low compactification scales, $M_C$, taking $n=1$ is
unrealistic since it implies large corrections to the Newton's law at solar distances.
Consequently, given the current experimental bounds~\cite{Long:1998dk}~\footnote{For
related bounds due to Casimir forces~\cite{Bordag:2001qi}, see also~\cite{Mostepanenko:2001fx}.}
  we have to imply either that there must be more than
one large extra compact dimensions, or  that the compactification scale is much smaller
than a few microns. Nevertheless, from the last formula one can
see that corrections near and below the compactification scale become
substantially large   and cannot be ignored.

The  closed form  derived for the case of
one ($n=1$) extra compact dimension~\cite{Floratos:1999bv}~\footnote{See
also~\cite{Oikonomou:2008fc}.}, allows
 to determine the behavior of the corrected potential even
inside the compact extra dimensional space where $r<R_C$,
however, for $n>1$ the sum as  expressed in (\ref{pot})
cannot be performed.  Instead,
we may use the Jacobi transformation to express  the potential as follows
 \ba
 \Phi(r,\theta)&\propto&\frac{1}{(2\sqrt{\pi})^{n+3}}\,
 \int_0^{\infty}\,ds\,s^{-\frac{n+3}{2}}\,
 e^{-\frac{r^2}{4s}}\,\sum_{k=1}^n\,e^{-\frac{\theta_kR^2}{4s}}
 \nonumber\\
 &&\times \left(1+2\sum_{m_k}^{\infty}\,e^{-\frac{m_k\pi
 R}{s}}\,cosh\frac{m_k\theta_k\pi R^2}{s}\right)\cdot
 \label{pot2}
 \ea
In order to examine the behavior of the potential, we first assume
zero angles and perform the integration. For two extra dimensions
the potential can be cast in the form  $\Phi_{n=2}=\Phi_{n=1}+\Delta\Phi_{12}$ with~\cite{future}
\ba
\Delta\Phi_{12}&=&G_N\,\frac{M\,M'}{R}\,\sum_{n=-\infty}^{\infty}\,\frac{4}{\rho_n}\,\frac{d}{d\,\rho_n}
\sum_{l=1}^{\infty}K_{0}(2\pi\,l\,\rho_n)
\ea
where $\rho=\frac{r}{2\pi R}$ and $\rho_n^2=\rho^2+n^2$.
Numerical investigation shows that the quantity $\Phi_{n=1}$ is the main
contribution to $\Phi(r,0)$.  Thus in the quantum problem the approximation (for
the $\theta=0$ case)
 of the $\Phi_{n=2}$ potential with $ \Phi_{n=1}$ case is sufficient for our purposes and
  $\Delta\Phi_{12}$ can be ignored.

\section{\protect \bigskip The Schr\"{o}dinger equation in extra compactified dimensions}

In this section, we  seek solutions of the $(n+3)$-dimensional Schr\"{o}dinger equation with the modified gravitational potential $\Phi(r,\theta)$ discussed above
\ba
-\nabla^2_{n+3}\Psi+\frac{2}{a_G}\Phi(r,\theta)\Psi&=&-\eps\,\Psi
\ea
where $a_G$ has been defined in (\ref{GA}) while we have introduced the parameter
\ba
\eps&=&-\frac{2m_nE}{\hbar^2}\cdot
\ea
For definiteness, here, we have taken $m_n$  to be the neutron mass and $M_0$ in the
$a_G$ definition (\ref{GA}) to be some point like mass generating the potential.
 Our aim is to determine the wavefunctions and the energy levels in the presence of  the modified gravitational potential.
We will assume that the involved particles are neutral, so that  electromagnetic
potential terms, which would normally  overwhelm any other source, are not present.
In the subsequent analysis, we will concentrate on the case of one extra
dimension only and introduce into  Schr\"{o}dinger's  equation the potential
(\ref{pot}) for $n=1$.

We first observe that the modified gravitational
potential exhibits an obvious  $2\pi$ periodicity with respect to
the parameter $\theta$ of the internal  compact dimension. We find it useful
 to use an established transformation between the radial part of the
Schr\"{o}dinger's equation of angular momentum $l$  to that of the
isotropic oscillation in $2\,l+2$ dimensions~\cite{Bergmann}. This transformation,
after the decoupling of the radial from the $\theta$ dependence, will help us to
transform the problem into an equivalent system of coupled Hill-type equations of
 periodic potentials~\cite{future}.

To start with, we parameterize the modified gravitational potential in the presence of
compact extra dimensions   as follows
\ba
\frac{2m\Phi(r,\theta)}{\hbar ^{2}}&\equiv&-\frac{f\left( r,\theta\right) }{r}<0\  \cdot
\ea
In the case $n=1$, the function $f(r,\theta)$ is  derived from (\ref{Vn1}) to be
\ba
f\left( r,\theta\right) &=&\frac{g}{R}\left( 1+2\sum_{k=1}^{\infty }e^{-\frac{r}{R}%
k}\cos \left( k\theta \right) \right)\label{ffun}
\ea
where
\ba
g&\equiv&\frac{2M_0m_n^{2}}{M_{C}^{3}}\;=\;2\frac{R}{a_G}\ \cdot\label{defg}
\ea
It should be noted that the new parameter $g$ introduced in (\ref{ffun}) is dimensionless.

 Firstly, we should point out that the limit $\theta=0$ and $r\ra 0$, is singular. We may
 further clarify this point considering the case $n=1$, where the potential is
given by the closed formula  (\ref{pot01}). We observe that the expansion for
small $r$ leads to the singular potential $\Phi(r)\sim \frac{1}{r^2}$. This is
of course consistent with the fact that for distances much smaller that the compactification
scale $r\ll R$ the potential assumes the familiar power law behavior (\ref{smallr}).
For weak couplings, however, as it is the case for the gravitational constant, the treatment
of the $1/r^2$ potentials is quantum mechanically consistent.
The failure of the potentials with higher singularities to produce a ground state is a well known fact which has
 been extensively discussed in the literature~\cite{Case:1950an}.
In the quantum mechanical treatment, the wavefunctions oscillate rapidly at the origin
and there is no way  to define a ground state. Nevertheless, away
from the origin, a consistent description is still possible. This corresponds to looking at excited energy levels where the wavefunctions are less sensitive to the tower of KK states,
which probe distances  close to the origin.

The radial part of the Schr\"{o}dinger equation has the familiar three dimensional form while  the extra-dimensions dependence is
encoded only in the potential through the function $f(r)\equiv
f(r,0)$:
\begin{equation}
\frac{d^{2}\mathcal{R}}{dr^{2}}+\frac{2}{r}\frac{d\mathcal{R}}{dr}-\left( \eps-%
\frac{f\left( r\right) }{r}+\frac{l\left( l+1\right) }{r^{2}}\right)
\mathcal{R}=0\cdot
\end{equation}%
This equation is to be trusted for distances higher than the compactification
scale $r>R_C$.

We now apply  the transformation   $r\rightarrow \frac{z^2}{2\sqrt{\eps}}$ , $\mathcal{R}\left( r\right)
\rightarrow \frac{p(z)}{(4\eps)^{7/8}z^{3/2}}$  to get
\begin{equation}
\frac{d^{2}p}{dz^{2}}-\left[ z^{2}+\frac{\left( 1+4l\right) \left(
3+4l\right) }{4z^{2}}-\frac{2}{\sqrt{e}}f\left( \frac{z^{2}}{2\sqrt{\eps}}%
\right) \right] p=0  \label{3dcb}
\end{equation}%
with the new parameter  $z$ being  dimensionless. Upon defining a new parameter
$ a=\frac{1}{2}\left( 1+4l\right)$
it is observed that equation (\ref{3dcb}) is a generalized form of the radial part of the
Schr\"{o}dinger's equation of angular momentum $l$  to that of the
isotropic oscillation in $2\,l+2$ dimensions.
It is easy to check that for the Newton's potential ($f(r)=1$)
the energy levels are given by $\eps_n=\frac{g^2}{4\,R^2\,(n+l+1)^2}$
and the principal quantum number is $N=n+l+1$.

A general analysis for all values of $l$ will be considered elsewhere~\cite{future}. Here,
we will only consider the case $l=0$ which corresponds to the value  $a=1/2$, so that (\ref{3dcb}) reduces to
\begin{equation}
\frac{d^{2}p}{dz^{2}}-\left[ z^{2}+\frac{3}{4z^{2}}-\frac{2}{\sqrt{e}}%
f\left( \frac{z^{2}}{2\sqrt{\epsilon}}\right) \right] p=0\  \cdot
\label{s1}
\end{equation}%
For $f(r)=1$ the differential equation (\ref{s1}) is reduced to the known simple
case~\cite{Case:1950an}  whose solutions are given in terms of the Laguerre functions,
\begin{equation*}
u_{n,\frac{1}{2}}(z)=z^{\frac{3}{2}}e^{-\frac{1}{2}z^{2}}{}_1F_1\left(
-n,2,z^{2}\right) \ =\frac{1}{n+1}z^{\frac{3}{2}}e^{-\frac{1}{2}%
z^{2}}L_{n}^{1}\left( z^{2}\right)   \cdot
\end{equation*}%
 We now express
the solution  $p\left( z\right) $ of  (\ref{s1}) as a functional series
in the eigenfunction basis of the $f(r)=1$   equation, where the expansion coefficients are to be determined
\begin{equation}
p\left( z\right) =\sum_{n=0}^{\infty }c_{n}u_{n,\frac 12}\left( z\right) \
\cdot \label{exp1}
\end{equation}%
Using the orthogonality properties of $L_{n}^{1}\left(z^{2}\right)$, we get the condition
\begin{equation}
\int_{0}^{\infty }u_{n,\frac{1}{2}}\left( z\right) u_{m,\frac{1}{2}}(z)dz=\frac{%
\delta _{mn}}{2\left( n+1\right) }\ \cdot\label{Lag1}
\end{equation}%
 Substituting the series (\ref{exp1}) into (\ref{s1})
and multiplying with $u_{m,a}\left( z\right) $, integration over $z$
gives
\begin{equation*}
\sum_{n=0}^{\infty }c_{n}\left[ \delta _{mn}-\frac{1}{\sqrt{e}}%
\int_{0}^{\infty }u_{m,\frac{1}{2}}\left( z\right)
u_{n,\frac{1}{2}}\left( z\right) f\left(
\frac{z^{2}}{2\sqrt{\epsilon}}\right) dz\right] =0\  \cdot
\end{equation*}%
In order to have a solution, the determinant of the above equation
must vanish. This vanishing determines the energy eigenvalues as
well as the expansion coefficients $c_n$ in (\ref{exp1}).\footnote{ Notice
that in the case of $f(r)=1$ in particular, using (\ref{Lag1}), we
simply recover the Balmer formula.}

 It is now straightforward to consider effects introduced by adding one extra dimension.
 To this end, we include the
 second-order derivative in the Laplacian for the extra compact dimension
 $x_c=R\,\theta$, while we restore the $\theta$-dependence in the
 potential. We expand now $p\left( z,\theta \right) $ in the same eigenfunction
basis $u_{n,\frac 12}$, but in terms of $\theta$-dependent coefficients
$c_{n}\left( \theta \right)$:
\ba
p\left( z,\theta \right) =\sum_{n=0}^{\infty }c_{n}\left( \theta
\right) u_{n,\frac 12}\left( z\right)\cdot \label{pzq}
\ea
We substitute the series into the Schr\"odinger equation and
multiply with $u_{m,\frac 12}\left( z\right) $. Finally, we  integrate as previously
over $z$ and we end up with a system of coupled differential equations
for the $\theta$-variable dependent expansion coefficients $c_n$:
\begin{eqnarray}
\sum_{n=0}^{\infty }\frac{1}{R^{2}\eps}\frac{d^{2}c_{n}\left( \theta \right) }{%
d\theta ^{2}}\int_{0}^{\infty }u_{m,\frac{1}{2}}\left( z\right) u_{n,\frac{1%
}{2}}\left( z\right) z^{2}dz+\left[
2-\frac{2}{\sqrt{\eps}}\int_{0}^{\infty }u_{m,\frac{1}{2}}\left(
z\right) u_{n,\frac{1}{2}}\left( z\right) f\left( \frac{z^{2}}{2\sqrt{\eps}}%
,\theta \right) dz\right] c_{n}(\theta)=0\cdot\label{HillDE}
\end{eqnarray}%
For  the first integral over the $z$-variable we find \cite{Prudnikov}
\begin{equation*}
A_{mn}=\int_{0}^{\infty }u_{m,\frac{1}{2}}\left( z\right) u_{n,\frac{1}{2}%
}\left( z\right) z^{2}dz=\frac{\left( -1\right) ^{m+n}\sin \pi
\left( m-n\right) }{\pi \left( m-n\right) \left[ 1-\left( m-n\right)
^{2}\right] }\ \cdot
\end{equation*}%
For $m=n$, the integral is $A_{nn}=1$.
For the second integral we first introduce the dimensionless parameter
$\alpha=\frac{1}{2R\sqrt{\eps}}$
and define
\ba
\alpha_k&=&\frac{k}{2R\sqrt{\eps}}\equiv k\alpha,\; k=1,2,\dots\label{noscaleng}
\ea
Then, the integral involving the function $f$ is a sum over $k$ of
integrals of the form
\begin{equation*}
B_{mn}(\alpha_k)=\int_{0}^{\infty }u_{m,\frac{1}{2}}\left( z\right) u_{n,\frac{1}{2}%
}\left( z\right) e^{-\alpha_kz^{2}}dz=\int_{0}^{\infty }z^{3}e^{-\left(
1+a_k\right) z^{2}}{}_1F_1\left( -m,2,z^{2}\right) {}_1F_1\left(
-n,2,z^{2}\right) dz\  \cdot
\end{equation*}%
 Using the relevant formula from \cite{Gradshteyn}, the coefficients $B_{mn}$ are found to be
\begin{equation}
B_{mn}(\alpha_k) =\frac{\Gamma \left( m+n+2\right) }{2\left( n+1\right)
!\left( m+1\right) !}\frac{\alpha^{n+m}}{\left( \alpha+1\right)
^{m+n+2}}\:{}_2F_1\left(
-m,-n;-m-n-1,\frac{\alpha_k^{2}-1}{\alpha_k^{2}}\right) \ \cdot
\end{equation}%
Now, all the coefficients in  (\ref{HillDE}) are known thus, we have
transformed the original Schr\"{o}dinger equation into a system of
an infinite number of Hill-type coupled Differential Equations for the $c_n(\theta)$'s.

\section{The Energy Shifts}

Let us now turn to the differential system (\ref{HillDE}).
Our aim is to determine the energy shifts as well as the modified wavefunctions
due to the presence of the additional potential terms escorting   the unperturbed
 Newton's potential. Because the extra terms
lead to an infinite number of coupled differential equations,
we naturally expect that the shift of any energy level will depend on
the infinite tower of the energy levels of the unperturbed equation.
It is further expected that the individual energy levels due to Bloch's theorem will turn
to energy bands.

As we have already said, in this work we elaborate on the case $\theta =0$ where  we
expect the effects to be maximal. Substituting the relevant form for the gravitational function
$f(r,0)$ into (\ref{HillDE}), the expansion coefficients $c_{n}(0)$ satisfy
\begin{equation*}
\sum_{n=0}^{\infty }\left[ \delta _{mn}-\frac{1}{\sqrt{e}}\frac{g}{R}%
\int_{0}^{\infty }u_{m,\frac{1}{2}}\left( z\right)
u_{n,\frac{1}{2}}\left(
z\right) \left( 1+2\sum_{k=1}^{\infty }\,e^{-\alpha_kz^{2}}%
 \right)dz\right]  c_{n}(0)=0\  \cdot
\end{equation*}%
Performing the integrations while defining
\begin{equation*}
D_{mn}(\alpha)=2\sum_{k=1}^{\infty }B_{mn}( \alpha_{k}),
\end{equation*}%
it follows that the above system reduces into a  linear
system of equations
\begin{eqnarray}
\sum_{n=0}^{\infty }\left[ \left( R\sqrt{\eps}-\frac{g}{2\left(
n+1\right) }\right) \delta _{mn}-gD_{mn}(\alpha)\right]\,c_{n}(0)& =&0 \ \cdot\label{lincoup}
\end{eqnarray}%
In this simplified form, we can easily observe that if
the coefficients $D_{mn}$ are set equal to zero, we immediately
obtain the standard `Coulombic' energy levels
${\eps_n}=\frac{g^2}{4R^2}\frac{1}{(n+1)^2}$. Thus, our task is to find the
modifications implied by the presence of the $D_{mn}$ contributions.
Due to the non-diagonal form of the latter it can be easily deduced that any energy
level $\eps_n$ receives corrections from an infinite number of energy
levels. Practically, we  aim to find the dimensionless eigenvalue
$R\sqrt{\eps_n}$ in terms of the dimensionless coupling
$g=\frac{2M_0m_n^{2}}{M_{C}^{3}}=2\frac{R_C}{a_G}$ using a finite but adequately large
number $N$ of states in (\ref{lincoup}).
It should be noted that this truncation does not presume that the coupling $g$ in front
of the correction terms $D_{mn}$ is small. As a matter of fact,
 it is expected that higher $n$-states will contribute less, so a sufficiently large
  number $N$ in the sum (\ref{lincoup})  will lead to a stable result.  The problem
  then is transformed to a $N\times N$-matrix equation where we are seeking solutions
 for the eigenvectors $\vec c(\theta)=(c_1,c_2,\dots c_N)^T$ and their corresponding
 energy eigenvalues $\eps_n$.   To further proceed,
  we use the definition
(\ref{noscaleng}) to write
\begin{equation*}
\sum_{n=0}^{\infty }\left[ \left( \frac{1}{\alpha
}-\frac{g}{\left( n+1\right) }\right) \delta _{mn}-2gD_{mn}\left(
\alpha \right) \right]c_{n}(0) =0\  \ \cdot
\end{equation*}%
The energy levels are then given by the solutions of the equation
\begin{equation*}
\left \vert \left( \frac{1}{\alpha }-\frac{g}{\left( n+1\right)
}\right) \delta _{mn}-2gD_{mn}\left( \alpha \right) \right \vert =0\
\  \cdot
\end{equation*}%
As noted, the above method works, even if $g$ is not in the perturbative region.
 However, for our present investigation, let us assume that  $g$ is small enough
so that we can handle the quantities $B_{mn}$ perturbatively.
Considering that the perturbative term  implies small corrections to the unperturbed
eigenvalues, we seek solutions of the form
\begin{equation*}
\frac{1}{\alpha _{n}}=\frac{g}{\left( n+1\right)
}+\sum_{k=3}^{\infty }c_{k}^{n}~g^{k}\  \cdot
\end{equation*}%
Omitting the calculational details (see~\cite{future}),  we finally get the following result:
\begin{equation*}
\alpha _{n}g=n+1-\frac{\pi ^{2}}{3}g^{2}+4\zeta \left( 3\right) g^{3}+\frac{%
\pi ^{4}}{9\left( 1+n\right) }g^{4}\  \cdot
\end{equation*}%
\begin{figure}[h]
\centering
\includegraphics[scale=.85]{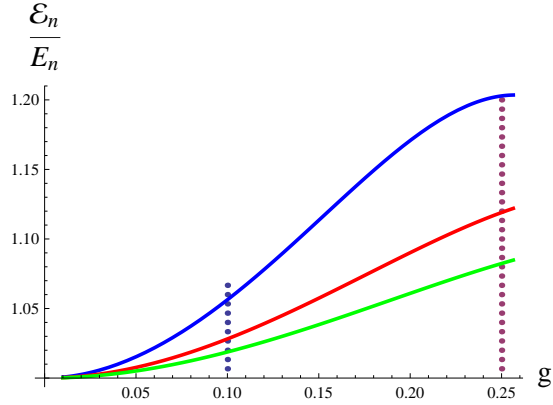}
\caption{(Color on line) The ratios  of the shifted energy levels over the Coulombic ones ${\cal E}_n/E_n$ for the first
three energy levels. The maximum deviation occurs at $g\sim \frac{1}{4}$ for the first Energy level (blue curve). }
 \label{Fshift}
\end{figure}
It is interesting to note that each Coulombic energy level  is shifted by a
constant up to order $g^{3}$ since the first three expansion
coefficients do not depend on the chosen dimensionality of $D_{mn}.$   We can write
the perturbed energy levels ${\cal E}_n$ in terms of the Coulombic ones $E_n=\frac{g^2}{4R^2(n+1)^2}$ as follows
\ba
{\cal E}_n&=&  \frac{E_n}{\left(1-\frac{\pi ^{2}}{3 (n+1)}g^{2}+\frac{4\zeta \left( 3\right)}{n+1} g^{3}+\frac{
\pi ^{4}}{9( 1+n)^2}g^{4}\right)^2}\label{Eshift}
\ea
In figure \ref{Fshift} we have plotted the correction to the gravitational energy levels for the
first three values of the principal quantum number $n=0,1,2$ and $l=0$. We observe that
for reasonable values of the coupling constant $g =2\frac{R_C}{a_G}$, between
$\left[\frac{1}{10}-\frac 14\right]$, the
corrections are experimentally detectable and are of the order of up to $15\%$. This means
that if the compactification radius is $R_C\approx \left[\frac{1}{10}-\frac{1}{100}\right]\,a_G$,
(where $a_G$ the `Bohr' radius of the gravitational atom), the corrections are sizable.
The existing experiments measuring the bound state energy spectrum of UCN beams on the  Earth's gravitational potential use specific geometries of horizontal systems of reflectors and absorbers and essentially measure the energy levels by the distance of the absorber of the reflector and the flux of the outgoing neutrons. The extra dimensional corrections to the Newton's potential we are
discussing here are negligible for this type of experiments. In our frame, we should
have experiments of UCN beams and spherically symmetric high-density materials  for
which a gravitational radius $a_G$ is larger than the compactification radius $R_C$ and the  radius of the
spherical material  $r_M$ should lie between these two:
\ba
R_C< r_M< a_G \ \cdot\label{ineq}
\ea
 To find if such materials exist in nature, we first recall the formula of the gravitational radius $a_G$ of a spherical object of mass $M_0$,
given in (\ref{GA}). Using the numerical values of the universal constants we express the radius $a_G$ in millimeters:
\ba
a_G&=&59.4\,\frac{1}{M_0/{\rm gr}}\,{\rm mm}
\ea
In terms of the material density $\rho_M$ and the radius of the spherical object inducing the gravitational
potential, the radius $a_G$ can be written
\ba
a_G&=& \frac{\hbar^2}{G_Nm_n^2}\frac{3}{4\pi \rho_Mr^3_M} \ \cdot\label{aGn}
\ea
Introducing the constant
\ba
\kappa&=& \frac{3\hbar^2}{4\pi G_Nm_n^2}=14.824\,{\rm gr\,mm}
\ea
the inequality (\ref{ineq}) gives the constraint for the density and the radius of the material
\ba
\frac{g}{2}\;<\;\frac{\rho_M\,r_M^4}{\kappa}\;<\;1\ \cdot\label{Constraint}
\ea
We can put the above constraint is a more useful form
\ba
\left(\frac{g}{2}\right)^4\,\frac{\kappa}{R_C^4}\,<\,\rho_M\,<\,\frac{g}{2}\,\frac{\kappa}{R_C^4}\
\cdot
\label{rhoRC}
\ea
For $g \sim 0.2$, using a spherical device of
the highest density material (Uranium), we can probe the extra dimension  down
to $R_C\sim 0.5$ mm.  If we decrease $g$, we can probe smaller distances, but the
perturbative energy shifts become tiny and rather hard to be detected by the experiment.

\begin{figure}[h]
\centering
\includegraphics[scale=.98]{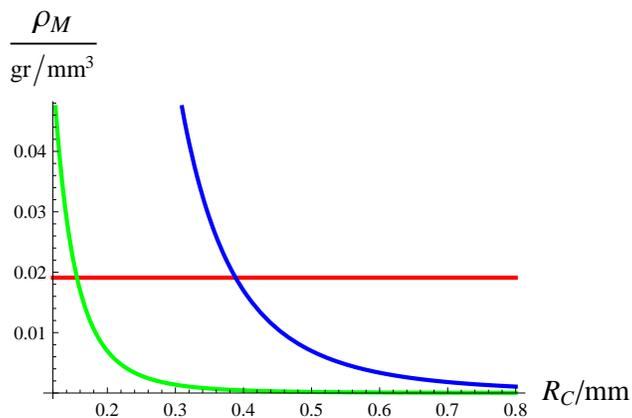}
\caption{(Color on line) Plot of the LHS limit of inequality (\ref{rhoRC}) for $g=\frac{1}{10}$ (lower curve) and $g=\frac 14$ (upper curve).
 The horizontal line corresponds to the density $19.1\times  10^{-3}$gr/mm$^3$ (Uranium).
  The corresponding energy shifts for the first three energy levels are found along the two vertical dotted lines in figure~\ref{Fshift}.}
 \label{MC}
\end{figure}

In figure~\ref{MC} we draw the left-hand  side (LHS) limit of the inequality (\ref{rhoRC}) in  the ($\rho_M,R_C$)-plane for two  characteristic values of $g=\frac{1}{10}, \,\frac 14$
 (for reasonable values of the expansion parameter $g$ the right-hand side
of (\ref{rhoRC}) is experimentally irrelevant).  For a given $g$-value
the region in the ($\rho_M,R_C$) plane for which the $R_C$ can be probed lies on the
right of the corresponding $g$-curve. For convenience, we have also plotted the
horizontal line $\rho_M=0.019$ gr/mm$^3$  which corresponds to the density of Uranium being the highest
density material existing in Nature. Thus the probed $R_C$'s correspond to the region
determined below this line and on the right to the $g$-curve.
We observe that for the existing densities in nature and reasonable $g$-values
the  compactification  radius $R_C$ is above  $\sim 0.2$ mm for the ground state
while present day experiments constrain $R_C$ to be smaller than $\sim 30$ microns for
 KK graviton scenarios of extra dimensions
which is our case. To probe smaller radii in our diagram we could consider capturing
the neutron into higher excited states $n=1,2,3,\dots$
One could think other geometries of the gravitational source so to probe smaller compactification
scales within the present gravity-modification scenario.

\section{Conclusions}

In this work we considered quantum gravitational effects produced by a modified gravitational potential from `decompactified' extra-dimensions with radii $R_C$ at the order of sub-micron scales.  We calculated the energy levels of a hypothesized `gravitational atom' formed by a neutron captured by a spherical mass. It was found that the energy-shifts $\Delta E_n$, compared to the energy  levels $E_n$ of the unperturbed $\frac 1r$-potential, can be expressed in terms of simple powers of the perturbative expansion parameter  $g=2\frac{R_C}{a_G}$ where $a_G$ is the `Bohr' radius of the `gravitational atom'.

We find that for reasonable values of the perturbative constant $g\sim [0.1-0.25]$, there are sizable $\Delta E_n\sim 10\%$ effects which are in principle measurable in properly designed experiments.  However, stringent limits on the size of extra dimensions require either smaller $g$-values where $\Delta E_n$
effects start becoming negligible,  or extremely dense  materials to generate a 'gravitational atom'  with sufficiently small $a_G$-radius.  Probes with the simple spherical geometry considered in this simple analysis are not sufficient to generate such small radii. We envisage that more sophisticated geometries
could be invented where these effects could be measured in future experimental explorations.

\vfill

{\bf Acknowledgements} The work of GKL and NDV is partially supported by the European Research and Training Network grant "Unification in the LHC era" (PITN-GA-2009-237920). The work of EGF is partially supported by the EKPA program Kapodistrias 70/4/9711.  GKL and EGF would like to thank the Physics Theory  group of \'Ecole Normale Sup\'erieure in Paris for kind hospitality during the last stage of this work.

\end{document}